\documentclass[preprint, showpacs, aps,amsmath]{revtex4}
\usepackage{amsthm}
\usepackage{amssymb}

\newcommand\revwidebegin{ \begin{widetext} }
\newcommand\revwideend{ \end{widetext} }
\usepackage{graphicx}

\theoremstyle{definition}

\theoremstyle{remark}

\newcommand{\E}{{\mathbb E}}
\newcommand{\F}{\mathcal F}
\begin{document}
\title[optimal WL algorithm]{ Optimal Modification Factor and Convergence of the Wang-Landau Algorithm}%
\author{Chenggang Zhou }%
\address{Quantitative Research, J.P. Morgan Chase \& Co., 12th Floor, 277 Park Ave. New York, NY 10017, U.S.A.}%
\author{Jia Su }
\address{Department of Chemistry, Princeton University, Princeton NJ, 08544, U.S.A.}
\email{Chenggang.X.Zhou@jpmorgan.com}%

\date{\today}%

\begin{abstract}
We propose a strategy to achieve the fastest convergence in the Wang-Landau algorithm with varying modification factors. With this strategy, the convergence of a simulation is at least as good as the conventional Monte Carlo algorithm, i.e. the statistical error vanishes as $1/\sqrt{t}$, where $t$ is a normalized time of the simulation. However, we also prove that the error cannot vanish faster than $1/t$. Our findings are consistent with the $1/t$ Wang-Landau algorithm discovered recently, and we argue that one needs external information in the simulation to beat the conventional Monte Carlo algorithm.
\end{abstract}
\maketitle

\section{Introduction}

Many researchers in various fields have been interested in using and improving the Wang-Landau(WL) algorithm.\cite{Wang01, Landau04}
This algorithm is in fact a general scheme to tackle problems that can be transformed into the problem of estimating a probability distribution.  This probability distribution can represent the simple density of states as a function of energy only,\cite{Wang01,  Landau04, Yamaguchi01, Okabe02, Brown05} but the more interesting and challenging applications are the joint density of states as a function of energy and a second variable.\cite{Shell02, Rathore02, Rathore03, Rathore04, Mastny05, Zhou06, Stefan07, Tsai07, Jutta08}
The second variable can be the volume of a liquid/gas system, the reaction coordinate of a polymer or a protein molecule, or the magnetization of a spin system. The (joint) density of states are subsequently used to calculate the partition function, but it is also doable to calculate the partition function directly with the WL algorithm without going through the density of states.\cite{Zhang07}
Apart from these physical models, several authors have explored the possibility of performing numerical integration with the WL algorithm.\cite{Troster05, Li07}  In contrast to alternative sampling methods, such as the umbrella sampling\cite{Torrie77} and multicanonical algorithm,\cite{Berg91} the WL algorithm almost seems to be a universal approach, because it does not rely on working out a good range of energy or distribution function to sample in advance.  However, most implementations of the WL algorithm involve some ``proprietary" enhancements designed to best suit the problem being studied. Some general strategies to improve the algorithm for a large class of problems have also been proposed.\cite{Zhou06, Yamaguchi02, Schulz02, Schulz03}

The original WL algorithm measures the histogram of the simulation, as soon as the histogram achieves a certain ``flatness", which is a tunable parameter in the simulation, the modification factor $f_n$ is reduced to $f_{n+1} = \sqrt{f_n}$. Many possible modifications have already been suggested and tested empirically.
One type of modifications involves the flatness criterion. Although to achieve a flat histogram is the initial motivation of the WL algorithm, Ref.~\onlinecite{Zhou05} suggested that the flatness is not a necessary criterion to achieve convergence.
Usually, the flatness is defined as the ratio of the fluctuation of the histogram (or other equivalent quantities) to the average histogram. In fact, fluctuation of the histogram is an intrinsic property of the WL algorithm.  A simulation with constant $f$ always improves the flatness, due to the increasing average histogram, but the amplitude of fluctuation eventually saturates.
The prediction in Ref.~\onlinecite{Zhou05} that the fluctuation of histogram is proportional to $1/\sqrt{\ln f}$ has been verified by different authors independently.\cite{Lee06, Morozov07}
Therefore, many authors often replace the flatness criterion with that of a minimum histogram. This also suggests that one should instead focus on the fluctuation of the histogram rather than the flatness. Actually the average histogram is absorbed into the normalization constant of the density of states, so the flatness is not a good indicator of convergence.

On the other hand, the exponential reduction of $f$ mainly came out of the expectation to achieve an exponential convergence.
Several authors have found the error of the original WL algorithm to saturate, while relaxing the exponential reduction rule seems to offer a better convergence and accuracy.\cite{Yan03, Jayasri05, Poulain06, Belardinelli07, Belardinelli08}  We mention in passing that this type of error saturation is different from the $\sqrt{\ln f}$ statistical error for a fixed $f$. Ref.~\onlinecite{Belardinelli07} and \onlinecite{Belardinelli08} clearly illustrate that even if $f$ is reduced to a very small value according to the original prescription, the statistical error stops to decrease at a certain point. This phenomenon suggests that exponentially reducing $f$ is not the best strategy.
Intuitive strategies to adjust $f$ have been proposed, such as increasing $f$ once in a while  during the simulation.\cite{Poulain06} The most impressive improvement has been observed for the simple $1/t$ rule.\cite{Belardinelli07, Belardinelli08} However a generic principle for achieving the optimal convergence is unknown.

The purpose of this paper is to derive such a principle to adjust $f$ in the simplest and most generic setting. The very first question one may ask is, what is the simplest and most generic setting of the WL algorithm? The literature so far is not clear on this issue. With implementation of improved algorithms mixed with unexplored models in different fields, the results are usually a convoluted mixture of model-specific properties and the generic behavior of the WL algorithm. One may suggest that the Ising model is a generic model to start with. However, it is easy to see that in terms of measuring the performance of the algorithm, the Ising model is not different from simply counting the number of states with the same total spin.\cite{Zhou05} If we flip a single spin in each update, the total energy or total spin can only change by a small amount every time. The WL algorithm compares $g(E_i)$ and $g(E_f)$, i.e. density of states before and after a move, but it does not require $E_i$ to be close to $E_f$. In principle, one can implement algorithms that allow "non-local" moves in the parameter space, e.g. cluster algorithm. The update schemes for the underlying model certainly have an effect on the outcome, which also makes rigorous analysis almost impossible. This is an effect that we want to factor out. So what models are more generic than the Ising model? Numerical integration\cite{Troster05, Li07,Belardinelli08} is a better candidate, since the evaluation of the integrand and moving from one sampling point to the other are both fast. The time spent in selecting the next sampling point is thus negligible.  Selecting the next sampling point $X$, may it be an energy, a 2-dimensional vector, or a value of the integrand, is an encapsulated process that depends on the actual problem to solve. What the high-level WL ``driver" asks for is only the next sampling point with probability distribution $P(X) \propto g(X)/\bar{g}(X)$, where $g(X)$ and $\bar{g}(X)$ are the exact and estimated density of states respectively.

This is in fact the starting point of analysis in Ref.~\onlinecite{Zhou05}, which mainly dealt with the dynamics of the simulation with a fixed $f$. This fluctuation leads to a statistical error in the density of states proportional to $\sqrt{\ln f}$. The multistage WL algorithm partially reduces this statistical fluctuation by decreasing $f$. However, its overall efficiency in reducing statistical error is not necessarily superior than the conventional Monte Carlo algorithm.\cite{Belardinelli07,Belardinelli08} This is why Ref.~\onlinecite{Zhou05} suggests to average over multiple independent simulations with a single $f$.

In Sec.\ref{sec2} of this paper, following the same approach, we derive an optimal strategy for updating the modification factor; in Sec. \ref{sec3}, we derive the upper and lower bound of the convergence with this optimal strategy, and compare it to the very impressive $1/t$ WL algorithm; Sec. \ref{sec4} presents our conclusion and discussions. The method we use in our study is entirely different from the well-known argument based on detailed balance. In fact, detailed balance applies to simulations with a set of constant transition probabilities. Most sampling algorithms fall into this class, including all importance sampling algorithms.
The basic Metropolis algorithm\cite{Metropolis53} is indeed the ``driver" of them. However, the transition probability in the WL algorithm is continually updated, therefore detailed balance does not directly apply to it. What we focus on in this paper is the WL ``driver". Similar to detailed balance, its mathematical validity warrants many applications, although not all physical models with specific pitfalls are guaranteed to be solved.  Therefore we only demonstrate our strategy on the most popular testing case, i.e. 2-dimensional Ising model. We believe the strategy we found is simple to implement in any existing simulations with the WL algorithm, and that its performance is comparable to the $1/t$ WL algorithm.

\section{Optimal modification factor}
\label{sec2}
 The WL algorithm is not a Markov process, because the transition probability depends on the history of the simulation. It might be necessary to re-iterate the main idea of Ref.~\onlinecite{Zhou05}. Suppose there are $N$ bins (energies) in a generic simulation, the key random process to focus on is the probability distribution $p_i(t)$ of the next move, with $i = 1,\ldots, N$. If bin $l$ is picked, this probability distribution evolves as:
 \begin{equation}
    p^{(l)}_i(t+1) = { p_i(t) f^{-\delta_{il}} \over 1-p_l(t) + p_l(t)f^{-1}}.
 \end{equation}
The transition probability from $p(t)$ to $p^{(l)}(t+1)$ is just $p_l(t)$. The process $p(t)$ is indeed a Markov process. $p(t)$ lives as a vector inside an $N$-dimensional simplex, whose center is the uniform distribution $\bar{p}_l = 1/N$ for $1\leq l \leq N$. Convergence means that the estimated density of states forces $p(t)$ to approach $\bar{p}$.

In Ref.~\onlinecite{Zhou05}, the level of convergence is measured by the function:
\begin{equation}
\mu(t) = N\ln N + \sum_{i=1}^N \ln p_i(t).
\end{equation}
 If the simulation has reached the exact density of states, therefore, a ``flat" histogram, then $p(t) = \bar{p}$, and $\mu(t) = 0$. In a simulation, $\mu(t)$ always starts from a negative value and approaches $0$ from below. The expected increment in $\mu(t)$ is given by
\revwidebegin
\begin{equation}
\label{eq2.3}
\E \left[ \mu(t+1) - \mu(t) | \F_t \right]
= -\ln f + N \sum_{k=1}^N p_k(t) \ln {1 \over
1- p_k(t)(1-f^{-1})}.
\end{equation}
\revwideend
Here the ``filtration" $\mathcal F_t$ is a Borel algebra which represents the information available at time $t$. $p(t)$ and
$\mu(t)$ are supposed to be available at time $t$, the expectation is over $N$ different $\mu(t+1)$.
It was proved in Ref.~\onlinecite{Zhou05} that if the distance between $p(t)$ and $\bar{p}$  is sufficiently large, $\mu(t+1)$ increases on average. Thus $\bar{p}$ serves as an attraction center that pulls $p(t)$ towards it. However, this attraction has a short-range cut-off at a characteristic distance $\zeta(f) =\sqrt{[(1-f^{-1})^{-1}\ln f -1] /N}$. At the same time $\zeta(f)$ determines how much the estimated density of states fluctuates around the exact density of states. Obviously the amount of fluctuation is reduced by decreasing $f$ in the simulation.

In practice, a systematic error in the simulation exists, which is a function of $f$ and the auto-correlation between successive additions to the histogram. As observed in Ref.~\onlinecite{Zhou05}, this systematic error becomes smaller when either $f$  or the correlation decreases. The amount of correlation is reflected by the tunneling time,\cite{Berg91,Dayal04} and Ref.~\onlinecite{Trebst04} offers a systematic way to decrease the tunneling time for one-dimensional density of states. But in general, decreasing the correlation (tunneling time) in the WL algorithm, especially for those calculating 2-dimensional joint density of sates, is not an easy task. This is why it is necessary to reduce $f$ to achieve accurate results. Otherwise, averaging over independent simulations as suggested in Ref.~\onlinecite{Zhou05} would be the best solution. In a multistage simulation, one has the freedom to chose $f_{n+1}$. We realize that there is an optimal $f$ in the sense that it maximizes Eq.~(\ref{eq2.3}). Selecting this optimal value for $f_{n+1}$ will bring $\mu(t)$ to zero at the fastest speed. An intuitive analogy is the random deposition model, where we would relate $f$ to the size of the next particle to be deposited. If $f$ is too small, it takes many such particles to fill up the existing depressions in the landscape. If $f$ is too large, instead of filling up the dips, the new depositions actually create a more hilly landscape.

To find this optimal $f$, we define $y = 1-f^{-1}$ and rewrite  the expected change Eq.~(\ref{eq2.3}) as
\begin{equation}
G(p(t), y) 
= \ln (1-y) + N \sum_{k=1}^N p_k(t) \ln {1 \over 1- p_k(t) y}.
\end{equation}
The Taylor expansion of this function
\begin{equation}
G(p,y) = N \sum_{n=1}^{+\infty} {1 \over n} \left( \sum_{k=1}^N p_k^{n+1} -{1\over N} \right) y^n,
\end{equation}
is easier to analyze. The coefficient of the first linear term is equal to $N \|p_k - \bar{p}\|^2$.
It is positive except when the simulation has perfectly converged. The higher order terms are negative if
$p_k(t)$ is in the vicinity of $\bar{p}$, which is the region that we care about most. In fact, when $p(t) \approx 1/N $ in this region,  the coefficient of $n$th term is roughly $ (N^{-n+1}-1)/n$, which is negative for $n>1$. (Assuming $N$ is a large integer.) The function $G(p,y)$ thus has a unique maximum value for $y \in [0,1)$. Suppose this
maximum value is achieved at $y_c$, then the corresponding $f_c = 1/(1-y_c)$ is the optimal modification factor that brings $\mu(t)$ towards zero at the maximum speed.

In the refining stage of an actual simulation, $y$ is close to zero, since $f$ is close to 1. In the following we will also refer to $y$ as the modification factor for convenience, as it is unlikely to confuse it with $f$. It should be
sufficient to retain two terms of the function $G(p, y)$:
\begin{equation}
    G(p, y) \approx N \|p -\bar{p}\|^2 y - {1\over 2}\left( 1 - N\sum_{k=1}^N p_k^3 \right) y^2 .
\end{equation}
With this approximation, the optimal value for $y$ is obtained as
\begin{equation}
\label{eq2.6}
y_c = {N \|p(t) - \bar{p} \|^2  \over 1 - N \sum_{k=1}^N p_k^3(t) }.
\end{equation}
On the denominator, since $p_k(t)$ is of order $1/N$ in the cases that we are interested in, the second term is of order $1/N$. For large systems, this correction can be safely omitted as well.

The nominator in Eq.~(\ref{eq2.6}) is simply a measure of the fluctuation of the estimated density of states, which is in agreement with our earlier analogy to the random deposition model. ``The optimal size of the deposition", i.e. $f_c \approx 1/(1 - N \|p(t)-\bar{p}\|^2)$, is determined by the current roughness of the landscape.

We illustrate this optimal strategy with the traditional testing case, the 2-dimensional Ising model. As suggested in Ref.~\onlinecite{Zhou05}, we update the
estimated density of states once every $K$ flips, so that the random walker has a chance to travel around between two updates in the estimated density of states. We arbitrarily set $K = N/2$ and the initial modification factor $\ln f_0 = 0.1$, but we have tested
that the simulation is stable with other reasonable choices.
Once all the energies have been visited, we start updating the modification factor.
Since  the exact density of states is known, $p(t)$ can be directly calculated as $p_l(t) = Z^{-1} g(E_l) / \bar{g}(E_l)$, where $Z$ is a normalization constant. The modification factor is updated with
\begin{equation}
\label{eq2.7}
\ln f = \alpha N \sum_l (p_l(t) - 1/N )^2,
\end{equation}
where $0<\alpha<1$, and we have used $\alpha = 0.1$.
[Note that only the first order term in the Taylor expansion of Eq.~(\ref{eq2.7}) matters. This
choice is also convenient since in actual simulations only $\ln \bar{g}(E)$ is used.]
The simulation continues with the updated $f$ until all energies have been visited again. We again update $f$ with Eq.~(\ref{eq2.7}) and run until all energies
are visited. This cycle repeats until the desired accuracy in density of states is reached,
which is also reflected in vanishing $\ln f$.

The choice of $\alpha$ requires some justification.
The simulation of Ising model, auto-correlation between updates in $\bar{g}(E_l)$ indeed exists, leading to the systematic error as discussed in Ref.~\onlinecite{Zhou05}. If we look at the error $\ln \left(\bar{g}(E) / g(E)\right)$,
we see the same slow-varying component in different simulations even the random number
seeds are different. The high-frequency fluctuations are indeed random.
However, in our simplified case for analytical analysis, all the
bins are equivalent and independent, which leads us to Eq.~(\ref{eq2.6}).
If we simply apply Eq.~(\ref{eq2.6}), $\|p(t)-\bar{p}\|^2$ contains contributions from
both this systematic error and the uncorrelated statistical fluctuation. As a result,
convergence is not guaranteed. Ideally one should
extract the uncorrelated statistical fluctuation, which can be estimated from the
high-frequency fluctuations of $\bar{g}(E)$, assuming $g(E)$ is a sufficiently smooth
function. However, the easiest fix that we have discovered is to insert a small positive
number $\alpha$ in Eq.~(\ref{eq2.7}).

On a second thought, one notices that the systematic error is brought into $\|p(t)-\bar{p}\|^2$
only because the exact density of states is known and used in our test. If it is not
known in advance, only the uncorrelated statistical fluctuation can be properly estimated
from $\bar{g}(E)$. One can only hope the systematic error to vanish as $f$ decreases, or
identify it with independent means. To guard against possible overestimation of the fluctuation,
we suggest to use $\alpha$ as an tunable parameter in all simulations. Thus, there are
only three configuration parameters, $f_0$, $K$, and $\alpha$. $f_0$ is a  trivial
one; both $K$ and $\alpha$ reduce the systematic error from auto-correlation in the sampling.
After setting these parameters, the simulations is worry-free. If the auto-correlation
turns out to be a problem, one can either increase $K$ or reduce $\alpha$ to fix it.

\begin{figure}[htbp]
\begin{center}
\includegraphics[width=0.85\columnwidth, height =0.6\columnwidth]{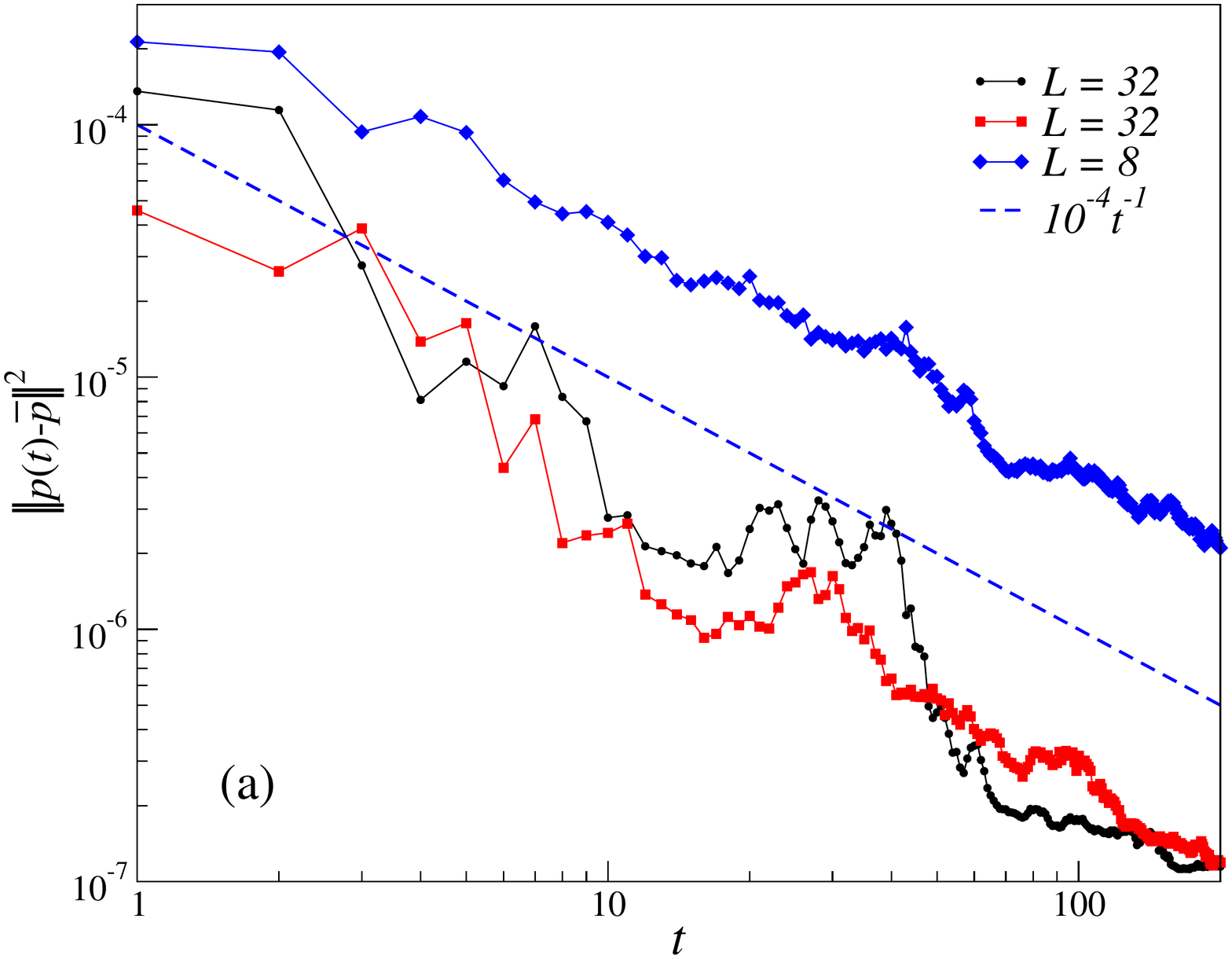}
\includegraphics[width=0.85\columnwidth,height = 0.6\columnwidth]{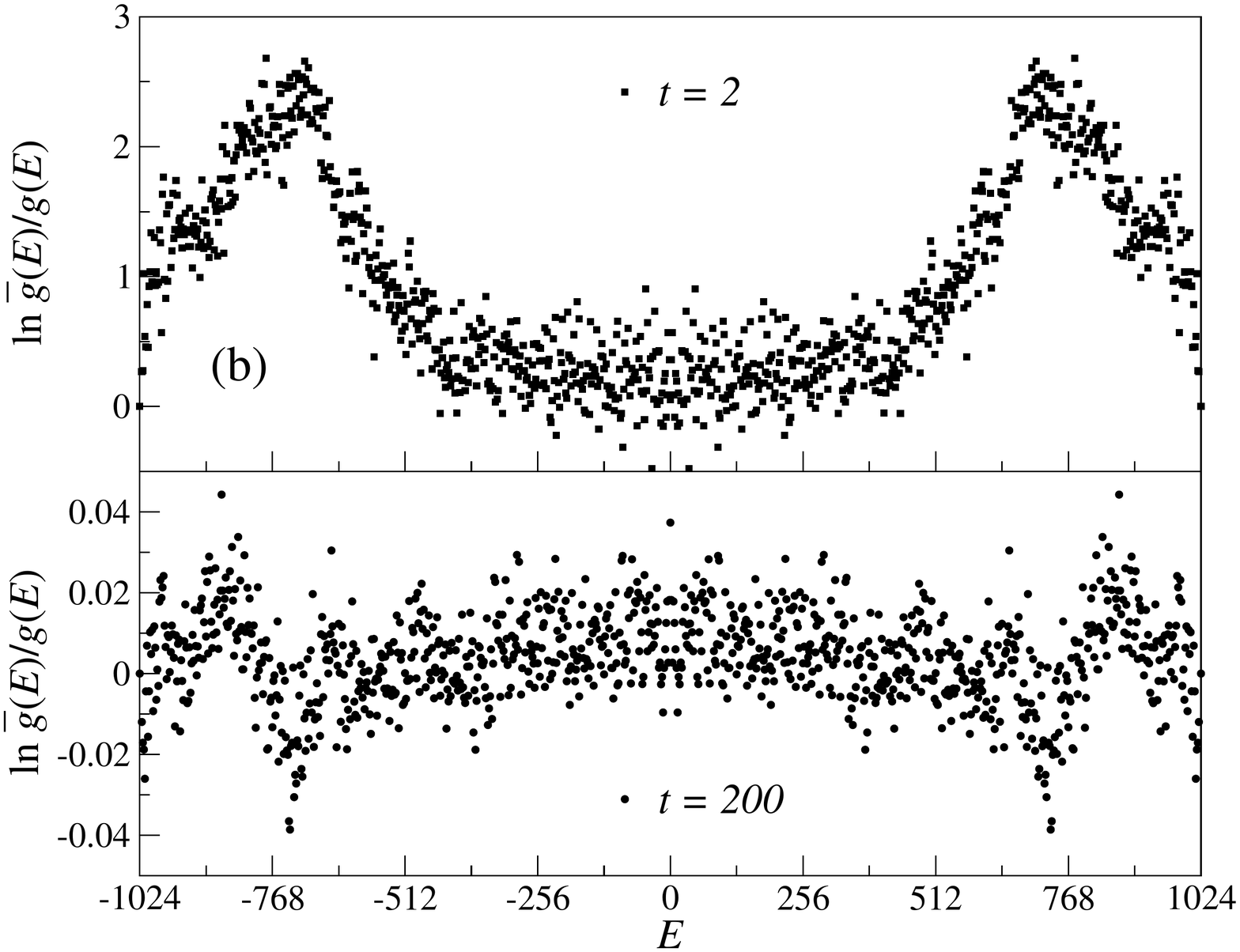}
\caption{Test with 2-dimensional Ising model. (a) (color online) The reduction of fluctuation, two runs of $L=32$ with
different random number seeds are presented. The fluctuation decreases as $1/t$ in the long run. (b) Comparison of error in
estimated density of states at different times. The systematic error is clearly visible in the upper panel at $t=2$, while
in the lower panel at $t=200$, the error is dominated by uncorrelated statistical fluctuation. }
\label{fig0}
\end{center}
\end{figure}

Figure~\ref{fig0}(a) plots the vanishing fluctuation observed in several simulations and the typical
pattern of error. These simulations were performed on square lattices with periodic boundary conditions. The linear size $L$ that we used ranges from $L=8$ to $L=32$. The exact density of states are calculated with the algorithm in Ref.~\onlinecite{Beale96}.
The fluctuation in $p(t)$ generally vanish as $1/t$, where $t$ is the number of cycles.
Since the simulation visits every energy at least once in each cycle, $t$ is approximately proportional to the total number of steps in the entire simulation. The $f$ in these simulations
also decreases roughly as $1/t$ since it is proportional to the fluctuation.
Figure~\ref{fig0}(b) shows that the typical shape of the error at the beginning of the simulation
and the final error. The initial error is shaped like a camelback, but its slow-varying
component is almost completely removed after 200 cycles.

A remaining question is how to estimate $p_k(t)$ and its distance to the uniform distribution, when the exact density of states is unknown. A simple approach is to
to keep a short histogram $H_i$ without modifying the estimated density of states, i.e.
setting $f=0$. Then $p_k(t)$ is estimated to be
\begin{equation}
    p_k(t) = { H_k \over \sum_{i=1}^N H_i }.
\end{equation}
Due to the discrete nature of the histogram, in order to estimate $p(t)$ to sufficient accuracy so that the calculated $y_c$ is not dominated by the sampling error of the histogram, the average height of the histogram has to grow larger as $p(t)$ approaches $\bar{p}$. This does not seem to be efficient in the long run.
One can also run multiple simulations with different random number seeds but the
same sequence of modification factors, the variation
in the estimated density of states is a measure of the statistical fluctuation
$\|p(t)-\bar{p}\|^2$. Anyway, estimating the statistical fluctuation can be a
time-consuming task in reality.

A better approach is to adjust $f$ according to a rule that is consistent with the natural dynamics of $\mu(t)$ governed by the optimal modification factor, so that extra simulation steps  to estimate  $\|p(t)-\bar{p}\|^2$ can be avoided. The fluctuation of $p(t)$ has $N-1$ degrees of freedom, where $N$ is usually a large integer, its distribution should resemble a narrow Gaussian peak centered on its mean value. Therefore, a natural choice is to replace $\|p(t)-\bar{p}\|^2$ in Eq.~(\ref{eq2.6}) with its expectation $\E [\|p(t)-\bar{p}\|^2]$, which we expect to depend only on the current $f$ and $t$.
In fact the test with Ising model suggests that the fluctuation converges roughly as $1/t$ and one may simply set
$\ln f_t = t^{-1}\ln f_0$, which is in agreement with the $1/t$ WL algorithm.\cite{Belardinelli07,Belardinelli08}
Why this power-law behavior appears and what are the possible values of its exponent will be investigated in the next section.

\section{Asymptotic behavior with the optimal $f$}
\label{sec3}

The instantaneous fluctuation $\|p(t)-\bar{p}\|^2$ is a random process. With the optimal modification factor, it is expected to converge in probability:
\begin{equation}
    \lim _{t\rightarrow +\infty} P( \|p(t)-\bar{p}\|^2 > \epsilon ) = 0, \forall \epsilon > 0.
\end{equation}
Due to the Markov inequality, it is a corollary  of
\[ \lim_{t\rightarrow +\infty} \E[\|p(t)-\bar{p}\|^2] = 0. \]
An upper bound for $\E[\|p(t)-\bar{p}\|^2]$ is sufficient to prove
this convergence, however, we need both tight upper and lower bounds to infer a useful rule of adjusting the modification factor.

\subsection{Upper bound}
\label{sec3.1}
For simplicity, we approximate $G(p,y)$ as \begin{equation}
G(p,y) = N\|p-\bar{p}\|^2y - {1\over 2}y^2.
\end{equation}
The omitted terms are of order $O(y^2/N)$. Once we adopt the optimal strategy, Eq.~(\ref{eq2.3}) becomes
\begin{equation}
    \E[\mu(t+1) | \F_t ] = \mu(t) + {1\over 2} N^2 \|p(t) - \bar{p}\|^4.
\end{equation}
It seems with the optimal modification factor, $\mu(t)$ still increases just by a small amount on average. We notice that in the vicinity of $\bar{p}$, i.e. $Np_k(t)-1 \ll 1$ ,
\begin{equation}
\mu(t) \approx - {1\over 2} N^2 \|p(t) - \bar{p}\|^2.
\end{equation}
Thus, let $S(t) = \|p(t+1) - \bar{p}\|^2$,  we can approximately write in this region:
\begin{equation}
\label{eq2.12}
    \E [ S(t+1)| \F_t ]
        = S(t) - S(t)^2.
\end{equation}
Both sides of this equation are $\F_t$-measurable random variables. One is tempted to replace the left side simply with $S(t+1)$, and hand-wavingly deduce that $S(t) \sim 1/t $ for large $t$. We approach more carefully here, since $S(t)$ is indeed a random process. Suppose we start with $S(0) \in (0, 1/2)$, the quantity that we are interested in is $\E [ S(t)| \F_0 ]$, which cannot be directly calculated with Eq.~(\ref{eq2.12}). To estimate its behavior for large $t$, we define an auxiliary sequence
\begin{eqnarray}
&&Z_0 = S(0),\\
&&Z_{t+1} = F(Z_t) = Z_t - Z_t^2,
\end{eqnarray}
and prove that $\E [ S(t)| \F_0 ] \leq Z_t$. Obviously, $Z_t > 0$ and $Z_{t+1}<Z_t$, for all $t = 0,1,\cdots$, therefore, $\lim_{n\rightarrow \infty} Z_t = 0$. It is also easy to see that $Z_t \sim 1/t$ as $t \rightarrow +\infty$, just by observing that
$Z_{t+1}^{-1} - Z_{t}^{-1} = 1/(1-Z_t)$, which monotonically converges to 1.
Next we use induction to prove $\E [ S(t)| \F_0 ] \leq Z_t$. This proposition is true for $t=0$. Suppose it is true up to $t=n$, for $t=n+1$, we have
\begin{eqnarray}
\nonumber
&&\E\left[S(n+1)| |\F_0 \right] =
\E\left[\E [ S(n+1)| \F_n ]|\F_0 \right] \\ \nonumber
&&\leq \E\left[S(n)|\F_0 \right]-\E\left[S(n)|\F_0 \right]^2 \\ \nonumber
&& \leq Z_n - Z_n^2 = Z_{n+1},
\end{eqnarray}
where we have used the Jensen inequality and the fact that $F(x)$ is increasing and concave in $(0,1/2)$. Thus, we have proved that the long term behavior of $\E [ S(t)| \F_0 ]$ is bounded by $1/t$.
If we assume that the error of a stochastic algorithm cannot vanish faster than $t^{-1/2}$, we can already conclude that $ \E [ S(t)| \F_0 ] \sim 1/t$, since the error in estimated density of states is proportional to $\sqrt{S(t)}$. Furthermore, the modification factor should decrease as $f_t \sim 1/t$.

\subsection{Lower bound}
Although now we have an upper bound for the asymptotic behavior of $\E [ S(t)| \F_0 ]$, we cannot rule out the possibility of an exponential decay. Consider a random process defined by $S_{t+1} =
\zeta_t(S_t - S_t^2)$, where $\zeta_t = 0$ or $2$ with probability $1/2$, and $S_0 \in (0,1/2)$. Then $P(S_n > 0) = 1/2^n$, and since the function $g(x) = 2(x-x^2)$ has a fix point at $1/2$, $\E [S_t | \F_0]$ decreases as $1/2^{t+1}$. Therefore, the distribution $P(S(t+1)|\F_t)$ has to meet certain requirements for the sequence $\E [ S(t)| \F_0 ]$ to decay as a power-law.

To rigorously prove that $\E [ S(t)| \F_0 ]$ does not decrease exponentially, our approach is to study the evolution of
a super-optimized version of the WL algorithm. At time $t$, if the $l$th bin is selected, the change in the measure $\mu(t)$ is
\begin{equation}
\label{eq3.7}
    \Delta \mu^{(l)}(t) = \ln (1-y) + N \ln { 1\over 1 - p_l(t) y }.
\end{equation}
Here we first apply the optimal modification factor $y = N\|p(t) - \bar{p}\|^2 = N S(t)$.
In order to make the fastest convergence, we can in principle let the computer choose the bin that gives the maximum increment in $\mu(t)$. One way to do this is to estimate $p(t)$ and choose the bin with largest $p_l(t)$ in every step, which is of course very difficult in reality. Actually, for models such as an Ising model, one cannot simply jump to the energy bin with the largest $p_l(t)$, but has to update the energy with a random walk. However, it is all right to perform this ``gedanken" simulation just for the sake of our analysis. In fact, when this choice is made, the simulation evolves deterministically.
\begin{eqnarray}
\nonumber
    &&\mu(t+1) - \mu(t) = \\
    && \max_{l} \left\{ \ln (1- NS(t) ) + N \ln { 1\over 1 - p_l(t) NS(t) } \right\} \\ \nonumber
    && \leq \ln(1-NS(t)) \\ \nonumber
    && - N \ln \left\{ 1 - \left[1/N+ \sqrt{(N -1)S(t)/N}\right] N S(t) \right\}.
\end{eqnarray}
The above inequality uses the fact that the maximum possible value for $p_l(t)$ subject to the constraint that $ \|p(t)-\bar{p}\|^2=S(t)$ and $\sum_l p_l(t) = 1$ is
$ 1/N+ \sqrt{(N -1)S(t)/N} $. Similar to the earlier derivation, we replace $\mu(t)$ with $-N^2S(t)/2$, and arrive at
\begin{equation}
\label{eq2.17}
 S(t+1) \geq S(t) - 2\sqrt{N-1\over N } S^{3/2}(t) - O(S(t)^2 N ^{-1}).
\end{equation}
Now assume the auxiliary sequence $Z_t$ is defined by $Z_{t+1} = Z_{t} - 2Z_{t}^{3/2}$, and $Z_0 = S(0)$. When $S(0)$ is sufficiently small so that the higher order terms
in Eq.~(\ref{eq2.17}) can be safely ignored, we can easily see that $S(t) \geq Z_t$.
On the other hand, $ \lim_{t\rightarrow \infty} Z_t t^2$ is finite, i.e. $Z_t \sim t^{-2}$. Clearly, even in the hypothetical super-optimized case, $S(t)$ is bounded from below by $1/t^2$, therefore, $\E [ S(t)| \F_0 ]$ can only vanish algebraically with an exponent in $[1,2)$. A hand-waving argument for the lack of exponential convergence is that in each step, the reduction of $S(t)$ never has a linear term in $S(t)$, but is always dominated by a power $S^\beta(t)$, with $\beta \geq 3/2$.

\begin{figure}[htbp]
\begin{center}
\includegraphics[width=0.9\columnwidth]{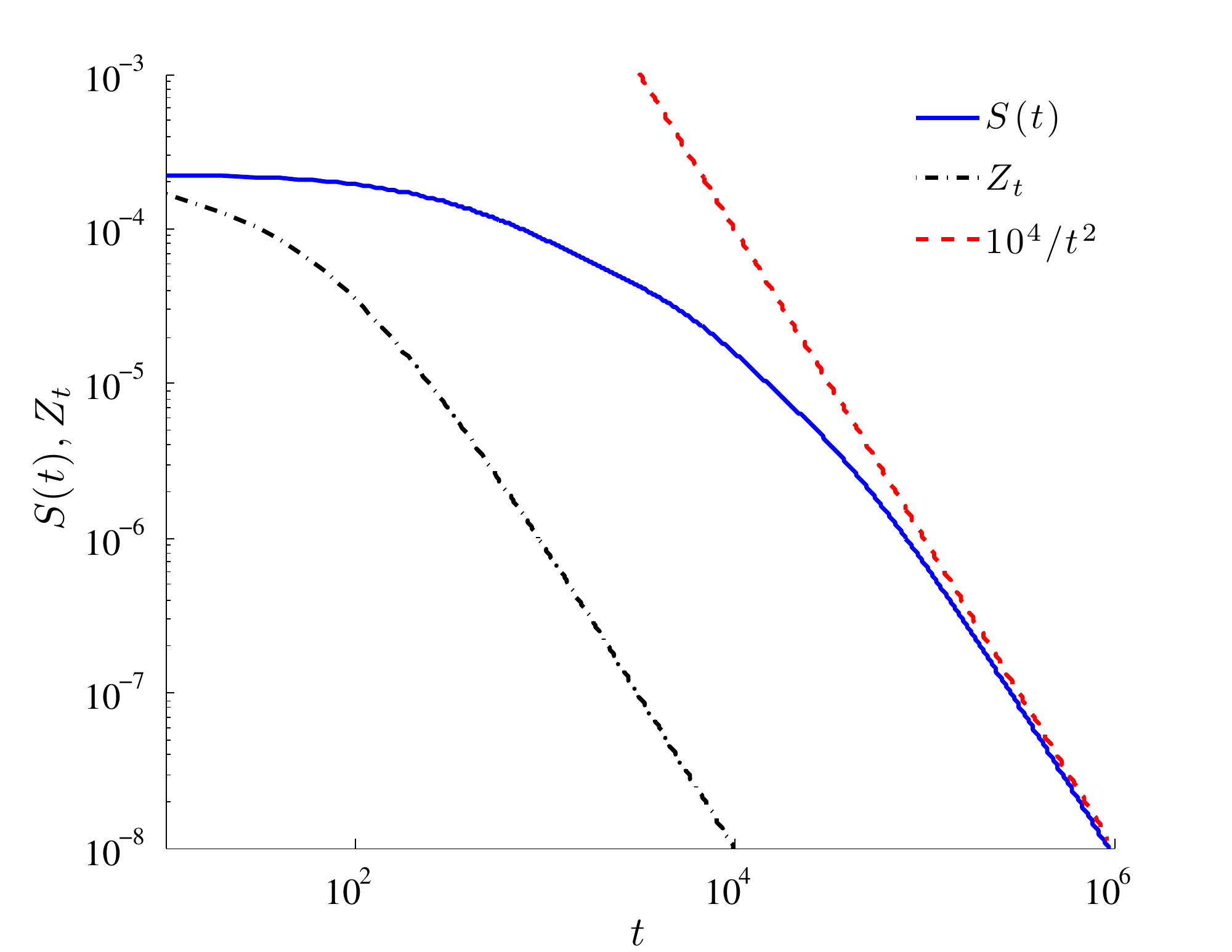}
\caption{Simulated sequence of $S(t)$ generated by the super-optimized algorithm and the auxiliary sequence $Z_t$. The dashed straight line is a guide for the eye, representing the $t^{-2}$ behavior.}
\label{fig1}
\end{center}
\end{figure}

Although this ``gedanken" simulation is not feasible for a meaningful physical model, it is trivial to numerically test the super-optimized algorithm with a small program. We have simulated the evolution of a vector $p(t)$ according to this deterministic evolution and the result is plotted in Fig.~\ref{fig1}. The initial value is set at $p_l(0) = 1/N + \epsilon_l$, where $\epsilon_l$ are small error terms whose sum is zero. The total error $S(t) = \|\epsilon(t)\|^2$ does vanish as $1/t^2$ asymptotically.

Note that the optimal $f$ applied above is based on the assumption that the next bin will be drawn randomly with probability $p(t)$. However, we eliminate the randomness after choosing $f$. With a second thought, if $p(t)$ is known, we can maximize Eq.~(\ref{eq3.7}) for both $y$ and $l$ and obtain
\begin{equation}
    \Delta \mu(t) \leq {1\over 2} N(N-1) S(t).
\end{equation}
The desired linear term $S(t)$ appears, which leads to an asymptotically exponential convergence. However, this ``ultra-optimized" algorithm is not feasible in actual simulation for the same reason given above. In fact, if we precisely know $p(t)$, there is no need to calculate it with a simulation. The uncertainty in the estimated $p(t)$ always results in an average over a certain probability distribution.

Now we have sufficient reason to believe that in the actual simulation $S(t)$ decreases as $1/t^\alpha$ with $1\leq \alpha < 2$. Correspondingly, the optimal modification factor is roughly given by $1 + NS(t)$, or $\ln f \approx NS(t) \sim 1/t^\alpha$. Since the actual simulation is stochastic, and there are other inefficiencies that we have ignored in the above analysis, it is likely to be safe to choose $\alpha = 1$. This choice is consistent with the recently discovered $1/t$ WL algorithm, which achieves an impressive $1/t^{1/2}$ convergence in error.

\subsection{Numerical simulation}

\begin{figure}[htbp]
\begin{center}
\includegraphics[width = 0.9\columnwidth]{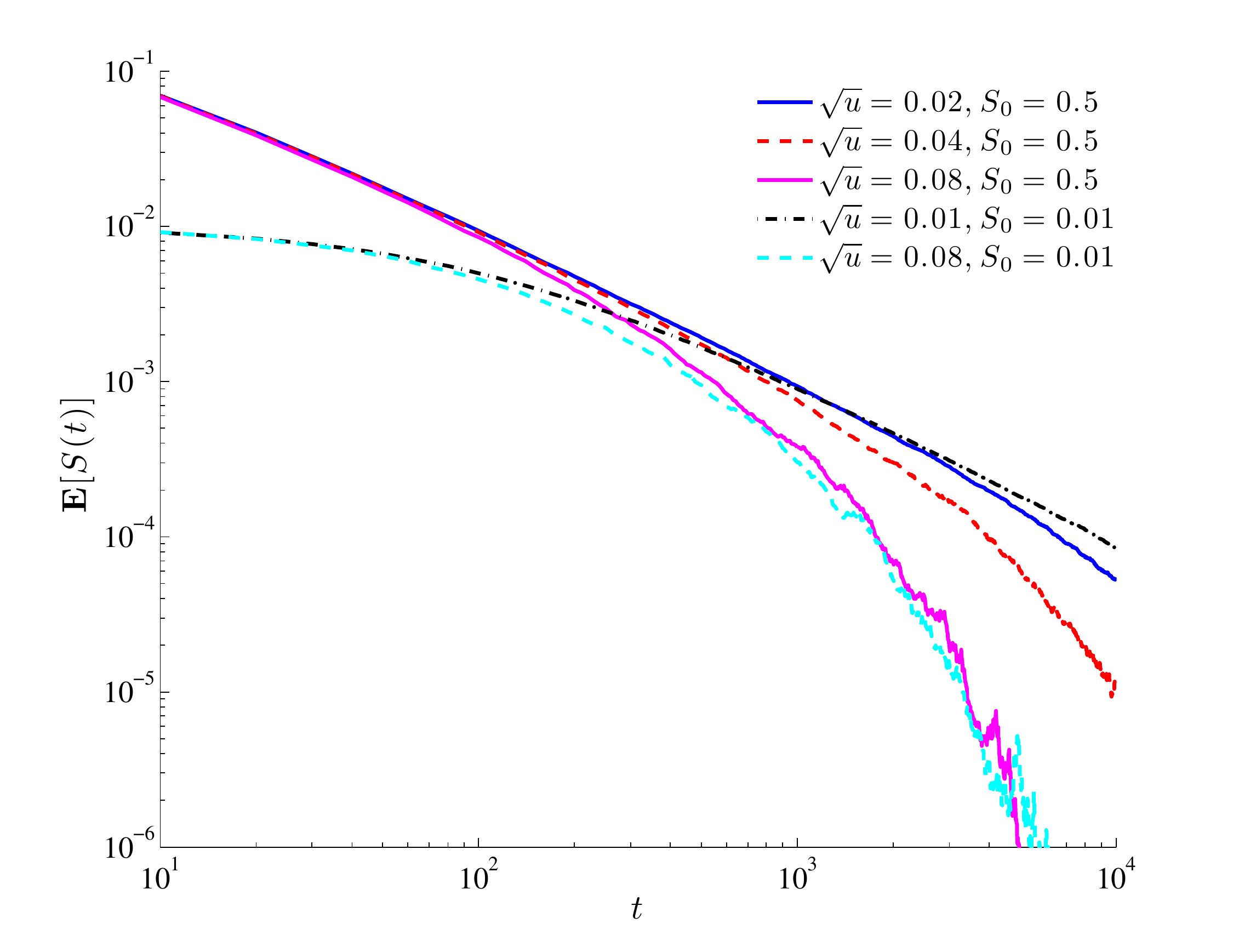}
\caption{(color online) Average trajectories of $S(t)$ with different initial condition $S(0)$ and the diffusion constant $u$. Every curve is the average of 1000 samples. These curves follows $1/(t+S_0^{-1})$ for small $t$, but crosses over to exponential decay at large $t$. }
\label{fig2}
\end{center}
\end{figure}

As we stated earlier, $S(t)$ is expected to be distributed in a narrow peak around its mean. We can write $S(t+1)$ as $S(t+1) = \zeta_t \left(S(t) - S^2(t)\right)$, where
$ \zeta_t$ is a random variable satisfying $\zeta_t \geq 0$ and $\E[\zeta_t] = 1$. We have given an example earlier that $\zeta_t$ can be chosen so that $\E[S(t)| \F ]$ decreases exponentially. It is interesting to ask the condition that $\zeta_t$ must satisfy so that $\E[S(t)| \F ]$ still decreases as $1/t$. We have performed numerical experiments to investigate the nonlinear stochastic evolution of $S(t)$, assuming $\zeta_t$ are independent lognormal random numbers, i.e.
$\zeta_t = e^{-u/2 + \sqrt{u} z}$, and $z$ is $N(0,1)$, or uniform random numbers with a box distribution, i.e. $P(\zeta) = a^{-1} {\bf 1}(1-a/2<\zeta_t<1+a/2)$.   Several averaged trajectories for $S(t)$ with different $u$ and $S(0) = S_0$ are plotted in Fig.~\ref{fig2}. They indicate that if the variance of $\zeta_t$ is small and $t$ is sufficiently small, then $\E[S(t)| \F ]$ decreases as $1/(t+ 1/S(0))$, but once $e^{ut} \gg 1$, $S(t)$ starts to decrease exponentially with large fluctuations. The simulations with box distribution for $\zeta_t$ exhibit the same behavior. These simulations are also numerical evidence of the $1/t$ upper bound that is proved in Sec.\ref{sec3.1}.

For small $u$ and $S_0$, we can approximate the stochastic difference equation of the process $S(t)$ with a continuous time stochastic differential equation:
\begin{equation}
    dS(t) = -S^2(t) + \sqrt{u} S(t) d B_t,
\end{equation}
where $B_t$ is the standard Brownian motion. This equation has an analytic strong solution:
\begin{equation}
\label{eq3.13}
 S(t) = {U(t) \over 1/S_0 + \int_0^t U(s) ds },
\end{equation}
where $S(0) = S_0$ and $U(t) = \exp\left( - ut/2 + \sqrt{u} B_t \right)$ is the standard exponential
martingale of the Brownian motion. The expectation value of $S(t)$ is not straightforward to calculate, but $\E[1/S(t)]$ can be quickly calculated using the properties of exponential martingale $U(t)$:
\begin{equation}
\label{eq3.14}
 \E\left[1 \over S(t)\right] = { e^{ut} \over S(0) } + {e^{ut}-1 \over u }.
\end{equation}
For small $t$ and $u$, by applying Taylor expansion to Eq.~(\ref{eq3.13}) and taking expectation term by term, we have obtained
\revwidebegin
\begin{eqnarray}
\label{eq3.15}
\E[S(t)] 
= S_0\left[1+ \sum_{n=1}^\infty (-)^n S_0^n \E \left[U(t)\int_0^t\cdots\int_0^t
 U(s_1) \cdots U(s_n) ds_1\cdots ds_n \right] \right]
= S_0 - S^2_0 \int_0^t {e^{us} \over (1+S_0s)^2}ds,
\end{eqnarray}
\revwideend
where we have also carelessly exchanged summation and expectation, and used the fact that $\E[ U(s_1) U(s_2) \cdots U(s_n)U(t)] =e^{us_n}$ when $s_1<s_2<\cdots s_n < t$. Obviously, this expression breaks down when the integral is larger than $1/S_0$. However, both Eqs.~(\ref{eq3.14}) and (\ref{eq3.15}) are consistent with the fact that the deterministic evolution $S(t) = S_0/(1+ S_0 t)$ is stable in the sense that for arbitrary $t$, one can find a $u$ small enough so that the error at time $t$ is smaller than any given amount. The noise controlled by $u$ makes $S(t)$ fluctuate around this deterministic evolution when $u$ and $t$ are relatively small, but when $e^{ut} \gg1$, the $S(t)$ becomes very noisy and $\E[S(t)]$ decreases exponentially.

Since in the actual simulations, the error has been observed to decrease as $1/\sqrt{t}$ seemingly forever, the above model with constant $u$ is probably not good enough for $t\gg 1/u$.  The effective $u$ in real simulation should decrease as $S(t)$ decreases.

\section{Conclusions and Discussions}
\label{sec4}
We suggest that the modification factor $f$ of the WL algorithm should be chosen to maximize the expected increment of the convergence measure $\mu(t)$, and that the resultant optimal modification factor is proportional to the fluctuation in the estimated density of states. With this optimal strategy, we have proved that the best behavior of the statistical error is bounded by $1/t^{1/2}$ and $1/t$, where the latter is deduced with the help of an ``gedanken" super-optimized WL algorithm. It is clear that the WL algorithm never converges exponentially, roughly speaking because the reduction in the fluctuation $S(t)$ is proportional to $S^\beta(t)$ with $ 3/2 < \beta \leq 2$. Since the optimal modification factor is proportional to $S(t)$, one should never decrease the modification factor exponentially.

Our estimation of the convergence is consistent with the recent numerical investigation of the $1/t$ WL algorithm.
We have proved that if the optimal modification factor is adopted, $\ln f \propto 1/t^{\alpha}$ with $1\leq \alpha < 2$. Actually, all existing numerical evidence suggest that in fact $ \alpha = 1$ is the best case for all pure Monte Carlo algorithms.
A test with the proposed strategy applied to Ising model also exhibits the $1/t$
convergence.
Numerical simulation of the process $S(t+1) = \zeta_t (S(t)- S^2(t))$ indicates that $\alpha = 1$ is stable when $\zeta_t$ has a small variance.
Our proof of the lower bound suggests that only if deterministic moves are mixed with the pure Monte Carlo algorithm, can a simulation achieve $\alpha > 1$.  Deterministic steps are not free, they need additional information about the density of states. Using a suitable initial estimation is one form of injecting information into the simulation. If new pieces of information are available during the simulation, it is possible to use them to modify the estimated density of states on the fly.
The fact that the super- and ultra-optimized deterministic algorithms have a faster convergence than the pure Monte Carlo algorithm is reminiscent of a wide variety of numerical integration algorithms, which converges at different speeds depending on the level of stochasticity in the algorithm.

Our results could have been a motivation for the $1/t$ WL algorithm. However, we have not proved the converse proposition that with the $1/t$ algorithm, the convergence in error is $1/\sqrt{t}$. Since our optimal $f$ is a random process that actually fluctuates around $1/t$, we expect the fluctuations to cancel out in the long run. Some approximations have been adopted in order to add to the readability of this paper. All the derivations presented here can be put into more rigorous formats. Finally, we stress that the focus in our study in this paper is the dynamic evolution of a random process extracted from the Monte Carlo simulation. This approach should be complementary to the traditional view that emphasizes on the stationary distribution and detailed balance of a Markov process. We hope this strategy and other techniques introduced here are of some general interest to research in Monte Carlo algorithms, especially those non-Markovian algorithms.\cite{Bussi06}

\bibliographystyle{amsplain}

\begin{thebibliography}{99}
\bibitem{Wang01}F. Wang and D. P. Landau, Phys. Rev. Lett. {\bf 86}, 2050 (2001); Phys. Rev. E {\bf 64}, 056101 (2001).

\bibitem{Landau04} D. P. Landau, S. Tsai, and M. Exler, Am. J. Phys. {\bf 72}, 1294 (2004).

\bibitem{Yamaguchi01}C. Yamaguchi, Y. Okabe, J. Phys. A {\bf 34}, 8781 (2001). 

\bibitem{Okabe02} Y. Okabe, Y. Tomita, and C. Yamaguchi, Compt. Phys. Commun. {\bf 146}, 63 (2002). 

\bibitem{Brown05} G. Brown and T. C. Schulthess, J. Appl. Phys. {\bf 97}, 10E303 (2005). 

\bibitem{Shell02}M. S. Shell, P. G. Debenedetti, and A. Z. Panagiotopoulos, Phys. Rev. E {\bf 66}, 056703 (2002). 

\bibitem{Rathore02} N. Rathore, J. J. de Pablo, J. Chem. Phys. {\bf 116}, 8745 (2002); ibid. {\bf 116} 7225 (2002). 

\bibitem{Rathore03} N. Rathore, T. A. Knotts IV, and J. J. de Pablo, J. Chem. Phys. {\bf 118}, 4285 (2003). 

\bibitem{Rathore04} N. Rathore, Q. Yan, and J. J. de Pablo, J. Chem. Phys. {\bf 120}, 5781 (2004). 

\bibitem{Mastny05} E. A. Mastny and J. J. de Pablo, J. Chem. Phys. {\bf 122}, 124109 (2005). 

\bibitem{Zhou06} C. Zhou, T. C. Schulthess, S. Torbr\"ugge, and D. P. Landau, Phys. Rev. Lett. {\bf 96}, 120201 (2006).

\bibitem{Stefan07} S. Torbr\"ugge and J\"urgen Schnack, Phys. Rev. B {\bf 75}, 054403 (2007). 

\bibitem{Tsai07}Shan-Ho Tsai, Fugao Wang, and D. P. Landau, Phys. Rev. E {\bf 75}, 061108 (2007). 

\bibitem{Jutta08} J. Luettmer-Strathmann, F. Rampf, W. Paul, and K. Binder, J. Chem. Phys. {\bf 128}, 064903 (2008). 

\bibitem{Zhang07} C. Zhang and J. Ma, Phys. Rev. E {\bf 76}, 036708 (2007). 

\bibitem{Troster05} A. Tr\"oster and C. Dellago, Phys. Rev. E {\bf 71}, 066705 (2005).

\bibitem{Li07} Y. W. Li, T. W\"ust, D. P. Landau, and H. Q. Lin, Comput. Phys. Commun. {\bf 177}, 524 (2007).

\bibitem{Torrie77} G. M. Torrie, J. P. Valleau, J. Comp. Phys. {\bf 23}, 187 (1977).
\bibitem{Berg91} B. A. Berg and T. Neuhaus, Phys. Lett. B {\bf 267}, 249 (1991); Phys. Rev. Lett. {\bf 68 }, 9 (1992).


\bibitem{Yamaguchi02} C. Yamaguchi, N. Kawashima, Phys. Rev. E {\bf 65}, 056710 (2002). 
\bibitem{Schulz02} B. J. Schulz, K. Binder, and M. M\"uller, Int. J. Mod. Phys. C {\bf 13}, 477 (2002). 
\bibitem{Schulz03}B. J. Schulz, K. Binder, M. M\"uller, and D. P. Landau, Phys. Rev. E {\bf 67}, 067102 (2003). 

\bibitem{Zhou05}C. Zhou and R. N. Bhatt, Phys. Rev. E {\bf 72}, 025701(R)(2005).

\bibitem{Lee06} H. K. Lee, Y. Okabe, and D. P. Landau, Comput. Phys. Commun. {\bf 175}, 36 (2006).
\bibitem{Morozov07}A. N. Morozov and S. H. Lin, Phys. Rev. E {\bf 76}, 026701 (2007). 


\bibitem{Yan03} Q. Yan and J. J. de Pablo, Phys. Rev. Lett. {\bf 90} 035701 (2003). 
\bibitem{Jayasri05}D. Jayasri, V. S. S. Sastry, and K. P. N. Murthy, Phys. Rev. E {\bf 72}, 036702 (2005).
\bibitem{Poulain06} P. Poulain, F. Calvo, R. Antoine, M. Broyer, and Ph. Dugourd, Phys. Rev. E. {\bf 73}, 056704 (2006).
\bibitem{Belardinelli07}R. E. Belardinelli and V. D. Pereyra, Phys. Rev. E {\bf 75}, 046701 (2007); J. Chem. Phys. {\bf 127}, 184105 (2007). 
\bibitem{Belardinelli08}R. E. Belardinelli, S. Manzi, and V. D. Pereyra, cond-mat/0806.0268. 

\bibitem{Metropolis53} N. Metropolis, A. W. Rosenbluth, M. N. Rosenbluth, A. M. Teller, and E. Teller, J. Chem Phys. {\bf 21}, 1087 (1953).
\bibitem{Dayal04}P. Dayal, S. Trebst, S. Wessel, D. Wurtz, M. Troyer, S. Sabhapandit, and S. N. Coppersmith, Phys. Rev. Lett. {\bf 92}, 097201 (2004). 
\bibitem{Trebst04}S. Trebst, D. A. Huse, and M. Troyer, Phys. Rev. E {\bf 70}, 046701 (2004). 

\bibitem{Beale96} P. D. Beale, Phys. Rev. Lett. {\bf 76}, 78 (1996). 

\bibitem{Bussi06} G. Bussi, A. Laio, and M. Parrinello, Phys. Rev. Lett. {\bf 96}, 090601 (2006).

\end{thebibliography}

\end{document}